# EFFICIENT OCT IMAGE SEGMENTATION USING NEURAL ARCHITECTURE SEARCH

*Saba Heidari Gheshlaghi [b,c], Omid Dehzangi[a,b,c], Ali Dabouei[c], Annahita Amireskandari[d], Ali Rezai[a,b], Nasser M Nasrabadi[c]*

[a]Department of Neuroscience, West Virginia University, USA
[b]Rockefeller Neuroscience Institute, West Virginia University, USA
[c]Lane Department of Computer Science and Electrical Engineering, West Virginia University, USA
[d]Ophthalmology and Visual Sciences, West Virginia University, USA
{sh0144@mix, omid.dehzangi@hsc, ad0046@mix, annahita.amireskandari@wvumedicine, ali.rezai@hsc, nasser.nasrabadi@mail}.wvu.edu

## ABSTRACT

In this work, we propose a Neural Architecture Search (NAS) for retinal layer segmentation in Optical Coherence Tomography (OCT) scans. We incorporate the Unet architecture in the NAS framework as its backbone for the segmentation of the retinal layers in our collected and pre-processed OCT image dataset. At the pre-processing stage, we conduct super resolution and image processing techniques on the raw OCT scans to improve the quality of the raw images. For our search strategy, different primitive operations are suggested to find the down- & up-sampling cell blocks, and the binary gate method is applied to make the search strategy practical for the task in hand. We empirically evaluated our method on our in-house OCT dataset. The experimental results demonstrate that the self-adapting NAS-Unet architecture substantially outperformed the competitive human-designed architecture by achieving 95.4% in mean Intersection over Union metric and 78.7% in Dice similarity coefficient.

## 1. INTRODUCTION

One of the common types of dementia is Alzheimer's disease (AD), in which aging is one of the most critical factors. Studies show that 5.8 million people in the US have AD, which means approximately 10% of the population older than 65 years old are suffering from this disease [1]. Currently, invasive medical procedures such as Position Emission Tomography (PET) and Magnetic Resonance Imaging (MRI) are employed for AD diagnosis. On the other hand, recent studies suggested that AD can be diagnosed and monitored by analyzing neuronal loss from the retina, which is related to the decrease of the retinal nerve fiber layer thickness [2][3]. Optical Coherence Tomography (OCT) is a non-invasive tool that produces cross-sectional images of the retinal structure. Therefore, it is hypothesized that OCT imaging providing the structure of the retinal layers that can be a useful biomarker in AD patients to evaluate and follow these patients in combination with the next generation of automatic machine learning allowing a detailed and yet efficient approach to quantify and identify the AD disease.

During the past decade, different CNN models (e.g. LeNet [4], GoogleNet [5], and DenseNet [6]) are utilized in different problems such as image segmentation, image recognition, speech recognition. The increasing demand in real life applications for developing new architectures with the highest efficiency and accuracy, which is very challenging and has been subjective (man-made architectures) given infinite possibilities. Moreover, finding the best hyper-parameters is significantly costly using human experts; therefore, there are many research efforts in the area of hyper-parameter tuning [7] such as random search [8] and Bayesian optimization [9]. However, it is still very challenging to find the best hyper-parameter set and model architecture for a predictive task in hand.

Recently, Neural Architecture Search (NAS), which is classified as a subfield of AutoML, plays a significant role in improving the network architecture to achieve higher accuracy without manual trial and error. NAS process has categorized into three steps: search space, search strategy, and performance estimation strategy. At the search strategy stage, NAS uses a recurrent neural network to sample different child architectures and train candidate architecture convergence in order to obtain their accuracies on the validation set. This accuracy is used (as a reward or punishment) to update the controller parameters; hence the controller will generate better architectures with higher accuracy over time [10]. However, NAS is very time-consuming and requires massive computational resources to achieve good results. Hence, new search strategies were presented to make the NAS more practical such as sharing parameters [11], searching from simple to complex [12], using Bayesian optimization [13], adaptively learning the best network [14], multi-level architecture encoding for parameter sharing [15], using gradient feedback from generic loss[16], or searching for architecture structure on smaller datasets and then generalizing for larger ones [17]. Moreover, other recent efforts have reduced the computation cost and made the NAS search more practical by concentrating more on finding the reputable cells and keep the backbone network fixed [18],[19], and [20].

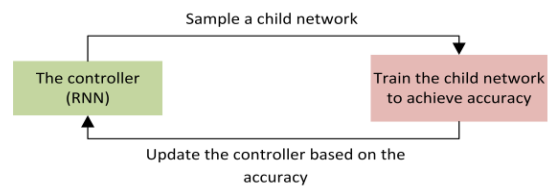

**Figure 1:** An overview of the neural architecture search.

## 2. RELATED WORK

In NAS algorithm development, most of the research is concentrated on image classification and are conducted on the CIFAR-10 dataset [21], [11], and [12]. Dense prediction cell [22], is one of the first attempts in using NAS for image segmentation by developing a recursive search space for dense image prediction with efficient random search. Also, the authors in [18] and [23] proposed the idea of continuous relaxation of a discrete optimization problem, which changes the search space from a discrete set of candidates to a continuous one. They used the gradient descent method on the validation set for optimization so that the controller part could be eliminated. This elimination helps the algorithm to be faster and more generalized in both recurrent and convolutional architectures. Auto-DeepLab [24] is a two-level hierarchical architecture search space that explores both cell and network structure. This method is a fundamental research in NAS image segmentation [25].

Medical image segmentation plays a significant role in the early diagnosis and treatment of diseases, and the Unet architecture [23] is a popular network in the field of medical image semantic segmentation. The Unet is an encoder-decoder network, in which the encoder part extracts the high-level context (semantic features), and the connection to the decoder part regenerates the spatial information and pixel classification results (image reconstructs) [26]. More recently, various Unet architectures provide better segmentation results and has become the most popular framework in medical image segmentation; hence, many researchers continue to work in this area and improve the algorithm using dense blocks to replace convolutional layers (both in the down- and up-sampling stages) [24], [25], [23], and [27]. As the demand growing in the AutoML area, the idea of combining NAS and Unet has been introduced [20]. The authors use the Unet like backbone and searches for the down- and up-sampling blocks by using the search strategy idea that was proposed in the Auto-Deeplab; however, this method eliminates the network structure search and only search for cell architectures (down- and up-sampling stages blocks) based on the U-like backbone. In this paper, we take the backbone architecture introduced in [20] as the backbone of our NAS-Unet algorithm and search for down- and up-sampling blocks that yield a higher accuracy on our in-house OCT medical image dataset.

## 3. METHODS

### 3.1. Data-sets:

We conduct the architecture search on our collected and pre-processed dataset comprising real OCT images. Our OCT dataset consists of 45 different subjects, each having 19 images collected at the Ophthalmology department at West Virginia University. These OCT images captured and unidentified (demanded by the IRB protocol) by a device made by the Heidelberg Engineering Company. Then, we labeled the data manually through the infinitt software, which is used by the clinicians to view patient reports and medical

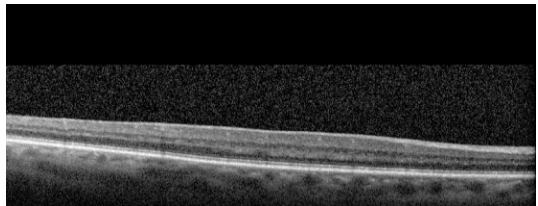

**Figure 2:** An example of a raw OCT image.

images. Then, we pre-processed the dataset as described next.

### 3.2. Pre-processing:

Improving the resolution and reducing the noise from input images is a crucial step in machine learning applications, which helps the model to converge faster and achieve better generalization performance. Super-resolution (SR) is one of the famous and effective techniques in enhancing image quality that takes a lower resolution input image and estimates a higher resolution version [28], [29], [30], [31], and [32]. In this article, the idea of super-resolution is applied as one of the pre-processing phases to have a higher image quality. Then, the mathematical morphological techniques such as erosion and closing are conducted to improve the image quality further. In our work, different super-resolution algorithms such as Enhanced Deep Residual Super-Resolution (EDSR) [29], Wide Deep Super Resolution (WDSR) [28], and Super Resolution GAN (SRGAN) [30] with scaling factor 4 are examined to generate the higher resolution OCT images before the modeling stage. These super resolution networks were previously trained and the final network has been used for the pre-processing step. OCT images that are applied to the SRGAN algorithm provide better enhancement results (as shown in Fig. 2) in comparison with other super-resolution algorithms (Fig. 3). Our NAS-Unet network is trained on images from the output of an SRGAN algorithm.

### 3.3. Search Strategy:

As mentioned above, U-like backbone architecture has two-part, the Down-Sampling Cell (DownSC) and Up-Sampling Cell (UpSC), which are going to be designed automatically by NAS. Inside each of these cells, the input nodes are the output of the two previous layers. In this work, the Conv-ReLU-GN order is assumed for all the convolutional operations (GN means group normalization) [33]. Table 1 indicates the operations that are used in searching cells. Cweight which is mentioned in table 1, states as squeeze-and-excitation. It is an operation that removes some redundant features by changing the feature map's dimensions to halve/double, before re-weighting channels.

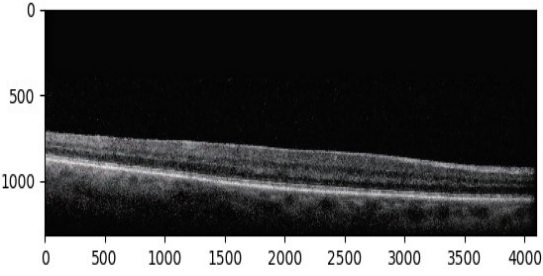

**Figure 3:** Generated image from SRGAN with Kernel size(5,5).

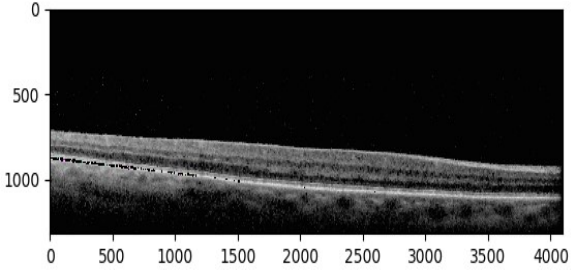

**Figure 2:** Generated image from WDSR with Kernel size(7,7).

**Table 1.** Type of primitive operation (PO) used for searching cell.

| DownSC operations | UpSC operations | Normal operations |
|---|---|---|
| Average pooling | Up cweight | identity |
| Max pooling | Up deph conv | cweight |
| Down conv | Up conv | conv |
| Down cweight | Up dilation conv | Dilation conv |
| Down dilation conv | - | Depth conv |
| Down depth conv | - | - |

In NAS research, the search space can be represented by using a single Directed Acyclic Graph (DAG). In a given neural network $C(e_1, ..., e_k)$, $e_i$ denotes an edge in the DAG, and $O = \{o_i\}$ is a set of the PO's as mentioned in Table 1. As shown in Fig. 4, all operations will be selected from Down/Up POs. The total number of edges between all intermediate nodes and input nodes are $K = 2M + M(M-1)/2$, where $M$ is the number of intermediate nodes (feature map layers).

Many articles and research works in the area of NAS search for the best search strategy process; however, searching directly for the architectures on large-scale tasks could be computationally expensive and challenging. In this work, to solve this issue and limitation, we use the idea of ProxylessNAS [19] that could directly learn the architectures for large-scale tasks. In the ProxylessNAS framework, a gradient-based approach for training the binarized parameters was presented, and, instead of fixing each edge to be definite from the primitive operation set, each edge could be characterized to be a mixed operation that has $N$ parallel paths, denoted as $m$. Hence, the over-parameterized network can be defined as $C(e_1 = m_O^1, ..., e_k = m_k^1)$. In ProxylessNAS framework with $x$ as an input, the output of a mixed operation is defined by the sum of $\{o_i(x)\}$ based on the output of its $N$ paths, while in DARTS [18], $m_O(x)$ is the weighted sum of $\{o_i(x)\}$, and for calculating the weights, softmax will apply to $N$ architecture parameters $\{\alpha_i\}$

$$\text{ProxylessNAS:} \quad m_O = \sum_{i=1}^{N} o_i(x). \quad (1)$$

$$\text{DARTS:} \quad m_O = \sum_{i=1}^{N} \frac{\exp(\alpha_i)}{\sum_j \exp(\alpha_i)} o_i(x). \quad (2)$$

The Eq. 1 & 2 calculate the $N$ feature map's output ($N$ is the number of candidate POs) and store them in the memory. However, training a compact model includes only one path. Hence, it needs $N$ times GPU for running, which could yield memory limitations. To solve this limitation, the idea of learning binarized path as a replacement for $N$ path was introduced in [19]. This idea consists of two alternative steps. First, for training the network and finding the weights, the architecture parameters are kept fixed and binary gates are sampled stochastically for each batch of input. Then, gradient descent is used for updating the weight parameters of active paths on the training dataset. At the next iteration, to learn the architecture parameters, the weight parameters are considered fixed, and the architecture parameters on the validation set are updated. Later, the final architecture can be extracted by removing the redundant paths after finishing training the architecture parameters.

## 4. EXPERIMENTS

In our investigation, we first applied image-processing techniques described in section 3.2 and particularly, SRGAN

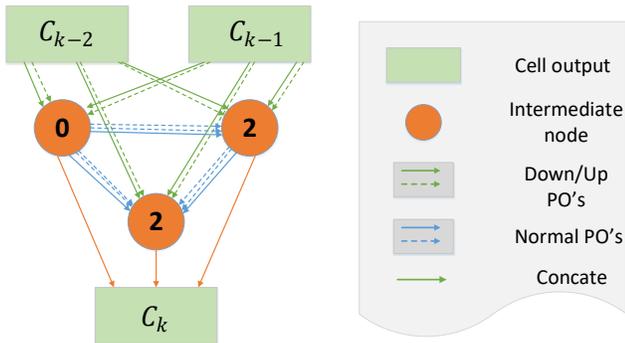

**Figure 5:** A cell architecture sample.

super resolution algorithm with kernel size of (5 × 5) is applied on our raw OCT scans to improve the resolution of and remove the noise from the images before the modeling (an example shown in Fig. 2). At the model training stage, we proposed a NAS-Unet algorithm to find the best neural network architecture with the highest performance in retinal layer segmentation on our OCT image dataset. In our experiments, we chose the Dice Similarity Coefficient (DSC) as the performance search criterion (i.e. minimizing the Dice loss). We set the number of intermediate nodes for both DownSC and UpSC blocks to 7, and our network searches over the primitive operations as listed in Table 1 to find the best DownSC and UpSC blocks leading to the highest DSC. We limit all the convolution operations to 3 × 3 size, and the pooling operations to 2 × 2 in order to have fewer parameters and spending less time to find the best architecture without any extra cost and/or sacrificing the network performance.

We divided our data into two mutually exclusive parts: 80% for train-set and 20% for test-set. Then, we conducted augmentation on our dataset by cropping our original 340 × 1024 OCT images to 300 × 300 images with overlapping pattern sliding from left to right through the middle of the image (i.e. 50% overlap), that is, 20 pixels are ignored from the top and bottom of the image. In addition, we applied horizontal flipping for further data augmentation required for the training of our proposed NAS architecture. At the search strategy stage, we use a binary gate update strategy with the batch size of 2, and the architecture search is conducted for 300 epochs. Moreover, SGD optimizer with momentum 0.95, cosine learning rate in the range of 0.025 to 0.01, and weight decay of 0.0003 has been set for learning the model weights. After learning the architecture, we iteratively improve our extracted Unet architecture with Adam optimizer on the OCT train-set. The initial learning rate and weight decay are set to 3.0e-4 and 5.0e-5, respectively. The entire architecture search process runs on two NVIDIA GeForce RTX 2080 Ti GPU having 12 GB RAM to accelerate training, and the search process takes about 1 day to run on the OCT scan dataset. Moreover, our network was implemented using Python 3.6.9, PyTorch 1.3.1, and scikit-learn 0.22.1 packages on Ubuntu 18.04.3 system.

## 5. RESULTS

To evaluate the performance and accuracy of our proposed NAS-Unet hybrid, both DSC and Mean Intersection over Union (mIoU) are reported. We evaluated the NAS-Unet performance by training it on OCT train-set, described in section 3.1, from scratch. We compare the Unet architecture derived from our proposed NAS architecture with human-designed Unet architecture by evaluating them on the OCT test-set. Table 2 reports the comparative results between the two architectures. Table 2 demonstrate that the Unet network derived by our NAS algorithm outperforms the human-designed model in all of the performance measures except in time complexity, which was expected. On the other hand, we decreased the computation time associated with our proposed NAS via efficient search strategy measures, described in section 3.3. Therefore, given the off-line nature of the segmentation task, our model training times are tolerable. Table 2 also reports that introducing SRGAN as a pre-processing step, had a good impact on our final performances (~16% & ~8% relative improvement on mIoU & DSC measures, respectively). In addition, the NAS-Unet pixel accuracy, mIou, and DSC results on our OCT dataset at different epochs are summarized chronologically, in Table 3. As shown in Table 3, our model consistently improved as the epochs developed and converged steadily.

**Table 2:** Comparing against conventional methods.

| Model | Dataset | mIoU | DSC | Time(Tr.) |
|---|---|---|---|---|
| Unet | OCT | 0.68 | 0.649 | **6h** |
| NAS-Unet without SRGAN | OCT | 0.945 | 0.769 | 1d-1h |
| **NAS-Unet with SRGAN** | **OCT** | **0.954** | **0.787** | 1d-4h |

**Table 3:** NAS-Unet network results on our OCT dataset.

| Epoch | Pixel Accuracy | mIoU | DSC |
|---|---|---|---|
| 50 | 0.610 | 0.597 | 0.492 |
| 100 | 0.692 | 0.672 | 0.521 |
| 150 | 0.853 | 0.850 | 0.595 |
| 200 | 0.915 | 0.913 | 0.659 |
| 250 | 0.937 | 0.935 | 0.785 |
| 300 | **0.955** | **0.954** | **0.787** |

## 6. CONCLUSION & FUTURE WORK

Optical Coherence Tomography (OCT) is a non-invasive, inexpensive, and timely efficient method that scans the human's retina with depth. It has been hypothesized that the thickness of the retinal layers extracted from OCTs could be an efficient and effective biomarker for early diagnosis of AD. In this work, we designed a self-training model architecture for the task of segmenting the retinal layers in real OCT scans. We fused the NAS algorithm with the Unet architecture as its backbone, which is the most popular network for medical image segmentation. At the pre-processing stage, we applied super resolution GAN and mathematical morphological techniques on our raw OCT dataset to achieve higher quality images before the modeling stage. In our architecture search strategy, 10 different primitive operations suggested to discover the best DownSC/ UpSC blocks. Moreover, binary gate method has been applied to save memory and make the search strategy more effective. The result of our search was evaluated by training on our dataset from scratch. The final derived NAS-Unet architecture significantly outperforms the human designed Unet network. For future work, we plan to expand our dataset to include sufficient normal vs. AD patients in order to discover distinguishing patterns including the layer thicknesses to develop an effective and efficient OCT-based AD diagnosis.